\documentclass[fleqn,usenatbib,useAMS]{mnras}
 \usepackage{graphicx}
 \usepackage{amsmath}
 \usepackage{amssymb}
 \usepackage[T1]{fontenc}
 \usepackage{ae,aecompl}
 \usepackage{txfonts}
 \usepackage{enumerate}
 \usepackage[hyphenbreaks]{breakurl}

 \title[On the origin of comet C/2018~V1]
       {Comet C/2018~V1 (Machholz-Fujikawa-Iwamoto): dislodged from
        the Oort Cloud or coming from interstellar space?
        }

 \author[C. de la Fuente Marcos and R. de la Fuente Marcos]
        {C.~de~la~Fuente~Marcos$^{1}$\thanks{E-mail: nbplanet@ucm.es}
         and
         R. de la Fuente Marcos$^{2}$ \\
         $^1$ Universidad Complutense de Madrid,
              Ciudad Universitaria, E-28040 Madrid, Spain \\
         $^2$AEGORA Research Group,
             Facultad de Ciencias Matem\'aticas,
             Universidad Complutense de Madrid,
             Ciudad Universitaria, E-28040 Madrid, Spain}
 \date{Accepted 2019 August 3.
       Received 2019 July 26;
       in original form 2019 February 17}
 \pubyear{2019}
 
\begin{document}
  \label{firstpage}
  \pagerange{\pageref{firstpage}--\pageref{lastpage}}
  \maketitle

  \begin{abstract}
     The chance discovery of the first interstellar minor body, 1I/2017~U1 
     (`Oumuamua), indicates that we may have been visited by such objects in 
     the past and that these events may repeat in the future. Unfortunately, 
     minor bodies following nearly parabolic or hyperbolic paths tend to 
     receive little attention: over 3/4 of those known have data-arcs shorter 
     than 30~d and, consistently, rather uncertain orbit determinations. This 
     fact suggests that we may have observed interstellar interlopers in the 
     past, but failed to recognize them as such due to insufficient data. 
     Early identification of promising candidates by using $N$-body 
     simulations may help in improving this situation, triggering follow-up 
     observations before they leave the Solar system. Here, we use this 
     technique to investigate the pre- and post-perihelion dynamical 
     evolution of the slightly hyperbolic comet C/2018~V1 
     (Machholz-Fujikawa-Iwamoto) to understand its origin and relevance 
     within the context of known parabolic and hyperbolic minor bodies. Based 
     on the available data, our calculations suggest that although C/2018~V1 
     may be a former member of the Oort Cloud, an origin beyond the Solar 
     system cannot be excluded. If extrasolar, it might have entered the 
     Solar system from interstellar space at low relative velocity with 
     respect to the Sun. The practical feasibility of this alternative 
     scenario has been assessed within the kinematic context of the stellar 
     neighbourhood of the Sun, using data from {\it Gaia} second data release, 
     and two robust solar sibling candidates have been identified. Our results 
     suggest that comets coming from interstellar space at low heliocentric 
     velocities may not be rare. 
  \end{abstract}

  \begin{keywords}
     methods: numerical -- methods: statistical -- celestial mechanics --
     comets: general -- 
     comets: individual: C/2018~V1 (Machholz-Fujikawa-Iwamoto) -- 
     Oort Cloud.
  \end{keywords}

  \section{Introduction}
     The first unambiguous detection of an interstellar minor body,\footnote{\url{https://minorplanetcenter.net//iau/mpec/K17/K17V17.html}} 
     1I/2017~U1 (`Oumuamua), on 2017 October 19 by R. Weryk \citep{2017MPEC....U..181B,2017MPEC....U..183M,2017MPEC....V...17W} is becoming 
     a puzzle in many respects (see e.g. the reviews by \citealt{2018Msngr.173...13H,2019NatAs...3..594O}), but also a game changer in the 
     study of how the Solar system interacts with interstellar debris. Finding this small object has opened a new window into our immediate 
     neighbourhood: material from beyond the Solar system may eventually be studied without having to resource to interstellar travel (see 
     e.g. \citealt{2018AJ....155..217S}). `Oumuamua had a hyperbolic excess velocity of about 26~km~s$^{-1}$ (see e.g. 
     \citealt{2017RNAAS...1a..21M}), but there must be interstellar comets and asteroids with lower hyperbolic excesses. 

     The distribution of the hyperbolic excess velocities of small bodies following hyperbolic paths with respect to the Solar system must 
     match that of stars observed in the solar neighbourhood as interstellar minor bodies must likely come from exoplanetary systems (see 
     e.g. \citealt{2018MNRAS.480.4903A,2018AJ....156..205B,2018MNRAS.479L..17P}). There are many examples of stars moving at low relative 
     velocity with respect to the Sun (see e.g. \citealt{2013AJ....145..102L}). On the other hand, \citet{1982ApJ...255..307V} have shown 
     that any interstellar minor body with a relative velocity not exceeding 0.5~km~s$^{-1}$ can be captured by the Solar system; in other 
     words, objects with relative velocity above 0.5~km~s$^{-1}$ can probably enter and leave the Solar system at will. This implies that 
     there should be a non-negligible number of interstellar interlopers with low hyperbolic excess velocities that were either missed 
     altogether by past and present surveys or have already been discovered but not yet identified as such (see a more general discussion in 
     \citealt{2018DPS....5020102A}). 

     De la Fuente Marcos, de la Fuente Marcos \& Aarseth (\citeyear{2018MNRAS.476L...1D}) have suggested that the Solar system might have 
     already been visited by interstellar comets, pointing out some suitable candidates to having an extrasolar provenance. However, it may 
     be argued that the Bayesian prior probability of a comet being interstellar must be rather low, given the fact that there are no 
     confirmed identifications of interstellar comets to date (other than `Oumuamua, if it is indeed a comet, see e.g. 
     \citealt{2018Natur.559..223M,2018ApJ...867L..17R,2019arXiv190108704S}). Within this context, it may not be justified to speculate about 
     past visits of interstellar comets on the basis of available data, but the fact is that, as of 2019 July 26, Jet Propulsion 
     Laboratory's (JPL) Small-Body Database (SBDB, \citealt{2011jsrs.conf...87G})\footnote{\url{https://ssd.jpl.nasa.gov/sbdb.cgi}} includes 
     2191 objects with nominal heliocentric eccentricity, $e\geq1$. Out of this sample, 531 objects have data-arcs longer than 30~d and 364 
     longer than 80~d (like the one of `Oumuamua). In other words, the probability of finding an interstellar minor body among those with 
     data-arcs as long as or longer than that of `Oumuamua is 0.00275; this estimate is probably lower than the one discussed by 
     \citet{2018ApJ...855L..10D}. Orbit determinations based on data-arcs shorter than one month are sometimes regarded as unreliable and 
     unsuitable to perform statistical analyses, but about 76 per cent of the known objects with $e\geq1$ fall into this category. If the 
     probability of finding an interstellar interloper (0.00275) is applied to the full sample (2191), six of them might have already been 
     discovered. These numbers suggest that there is a strong bias against detecting interstellar minor bodies, which may not be correctly 
     identified simply because they do not generate enough attention to be re-observed and their orbit determinations remain consistently 
     poor during their relatively short observability windows. 

     As shown by \citet{2018MNRAS.476L...1D}, $N$-body simulations may help in singling out promising candidates to be re-observed before 
     they leave the Solar system, never to return. Here, we present a detailed analysis of the pre- and post-perihelion dynamical evolution 
     of C/2018~V1 (Machholz-Fujikawa-Iwamoto) ---a slightly hyperbolic comet--- aimed at identifying the most probable provenance of this 
     object. This paper is organized as follows. In Section~2, we present the data and the tools used in this study, and discuss the 
     scientific case. Section~3 investigates the orbital context of C/2018~V1. The dynamical evolution of C/2018~V1 is explored in 
     Section~4. In Section~5 and as a practical exercise, we use data from the {\it Gaia} second data release (DR2) to investigate a 
     putative kinematic link to stars in the neighbourhood of the Sun. Our results are discussed in Section~6 and our conclusions summarized 
     in Section~7.

  \section{Data and methods}
     The source of most of the data used in this research is JPL's SBDB and JPL's \textsc{horizons}\footnote{\url{https://ssd.jpl.nasa.gov/?horizons}} 
     ephemeris system \citep{1996DPS....28.2504G}. This includes orbit determinations, covariance matrices, initial conditions (positions 
     and velocities in the barycentre of the Solar system) for planets and minor bodies referred to epoch JD 2458600.5 (2019 April 27.0) TDB 
     (Barycentric Dynamical Time) ---which is the zero instant of time in the figures, J2000.0 ecliptic and equinox, unless explicitly 
     stated--- ephemerides, and other input data. Another data source is in {\it Gaia} DR2 \citep{2016A&A...595A...1G,2018A&A...616A...1G} 
     that provides extensive astrometric and photometric data of stellar sources in the form of coordinates, $\alpha$ and $\delta$, 
     parallax, $\pi$, radial velocity, $V_{\rm r}$, proper motions, $\mu_{\alpha}$ and $\mu_{\delta}$, and their respective standard errors, 
     $\sigma_{\pi}$, $\sigma_{V_{\rm r}}$, $\sigma_{\mu_{\alpha}}$, and $\sigma_{\mu_{\delta}}$. Such data are used to perform a systematic 
     search for stars with kinematic properties consistent with those of the pre-encounter trajectory of C/2018~V1 (in the unbound case) as 
     discussed in Section~5.  

     \subsection{Comet C/2018~V1 (Machholz-Fujikawa-Iwamoto): data}
        Using a 0.47-m reflector, D.~E.~Machholz reported the visual discovery of a comet on 2018 November 7 (observing from Colfax, CA, 
        U.S.A.); the same object was independently discovered by S.~Fujikawa (observing from Kan'onji, Kagawa, Japan) and M.~Iwamoto 
        (observing from Awa, Tokushima, Japan), and assigned the temporary name 
        TCP~J12192806-0211143.\footnote{\burl{http://www.cbat.eps.harvard.edu/unconf/followups/J12192806-0211143.html}} The Minor Planet 
        Center (MPC) first listed the new object in its NEO Confirmation Page (NEOCP) as DM001 and then moved it to the Possible Comet 
        Confirmation Page (PCCP) before issuing three MPECs \citep{2018MPEC....W...46B,2018MPEC....W...03H,2018MPEC....V..151M} with 
        observations and orbit determinations under the new official name of C/2018~V1 (Machholz-Fujikawa-Iwamoto). Almost concurrently,
        the Central Bureau of Electronic Telegrams issued three CBETs \citep{2018CBET.4572....2G,2018CBET.4572....3N,2018CBET.4572....1S}. 

        The orbit determination of C/2018~V1 initially available from JPL's SBDB and computed by J.~D. Giorgini on 2018 November 20, see 
        Table~\ref{elements}, was based on 625 data points for a data-arc span of 12 days. The orbit of the comet was slightly hyperbolic, 
        for this initial determination, at the 1.5$\sigma$ level when considering the heliocentric values (but 27.6$\sigma$ when considering 
        the barycentric ones). The uncertainties in the orbit determination of C/2018~V1 were similar to those affecting the one of 
        1I/2017~U1 (`Oumuamua) several weeks after discovery. 

        On 2019 May 15, a second orbit determination was made public that included additional observations that had surfaced over the 
        previous months. The new orbit determination is shown in Table~\ref{elements} and it is based on 750 data points for a data-arc span 
        of 37 days. The orbit remains slightly hyperbolic, now at the 16.4$\sigma$ level when considering the heliocentric values (but 
        510.9$\sigma$ when considering the barycentric ones). The uncertainties are now comparable to those of the final orbit determination 
        of `Oumuamua; therefore, we are confident that the new orbit determination is robust enough to produce sufficiently reliable initial 
        conditions that can be used to make predictions regarding the past, present, and future orbital evolution of this interesting 
        object.
%
%
      \begin{table*}
         \centering
         \fontsize{8}{11pt}\selectfont
         \tabcolsep 0.25truecm
         \caption{\label{elements}Heliocentric and barycentric Keplerian orbital elements of comet C/2018~V1 (Machholz-Fujikawa-Iwamoto). 
                  Heliocentric values include the 1$\sigma$ uncertainty. The first orbit determination (2018 November 20) is referred to 
                  epoch JD 2458436.5, which corresponds to 0:00 on 2018 November 14 TDB, and it was produced by J.~D. Giorgini (J2000.0 
                  ecliptic and equinox). It is based on 625 observations with a data-arc span of 12 days. The second orbit determination 
                  (2019 May 15) is referred to epoch JD 2458438.5, which corresponds to 0:00 on 2018 November 16 TDB. It is based on 750 
                  observations with a data-arc span of 37 days. Source: JPL's SBDB.
                 }
         \begin{tabular}{lccccc}
            \hline
                                                               &   & \multicolumn{2}{c}{2018 November 20}  &  \multicolumn{2}{c}{2019 May 15}      \\
             Parameter                                         &   & heliocentric            & barycentric & heliocentric            & barycentric \\
            \hline
             Perihelion, $q$ (au)                              & = &   0.386970$\pm$0.000011 &   0.394399  &   0.386954$\pm$0.000002 &   0.394183  \\
             Eccentricity, $e$                                 & = &   1.0006$\pm$0.0004     &   1.0107    &   1.00040$\pm$0.00002   &   1.01235   \\
             Inclination, $i$ (\degr)                          & = & 143.989$\pm$0.002       & 144.206     & 143.9878$\pm$0.0005     & 144.2182    \\
             Longitude of the ascending node, $\Omega$ (\degr) & = & 128.724$\pm$0.006       & 128.400     & 128.7222$\pm$0.0004     & 128.4031    \\
             Argument of perihelion, $\omega$ (\degr)          & = &  88.769$\pm$0.007       &  87.840     &  88.7749$\pm$0.0002     &  87.8495    \\
            \hline
         \end{tabular}
      \end{table*}
%
%

     \subsection{Kinematic context: \textbf{\itshape{Gaia}} DR2 data}
        In Section~5, we use data from {\it Gaia} DR2 to investigate a putative kinematic link of C/2018~V1 to stars in the neighbourhood of 
        the Sun. In order to interpret our results properly (e.g. by using colour-magnitude diagrams or CMDs), we focus on those sources 
        with estimated values of the line-of-sight extinction $A_G$ and reddening $E(G_{\rm BP}-G_{\rm RP})$; {\it Gaia} DR2 includes 
        87\,733\,672 such sources,\footnote{\url{https://www.cosmos.esa.int/web/gaia/dr2}} all of them have strictly positive values of the 
        parallax. Out of this sample, 4\,831\,731 sources have positions, parallax, radial velocity, and proper motions. This smaller sample 
        is used to investigate a possible kinematic link of C/2018~V1 to neighbour stars. We have not performed any correction to address 
        the issue of the zero-point offset in {\it Gaia} DR2 parallax data. There are several independent determinations of its value (see 
        e.g. \citealt{2018ApJ...861..126R,2018ApJ...862...61S,2019AstL...45...10B,2019ApJ...875..114X,2019ApJ...878..136Z}) and most of them 
        are larger than the value of 0.029~mas initially presented by \citet{2018A&A...616A...2L}. The zero-point offset values quoted in 
        the recent literature range from 0.029 to 0.082~mas that must be added to the original values in {\it Gaia} DR2 in order to perform 
        the correction (i.e. in general the actual distances to the sources may be shorter than those computed directly from the catalogue
        data). Neglecting this correction has no significant effect on our conclusions as the uncertainties of interest here (see Section~5) 
        are larger or of the same order as the value of the correction. Additional details on our {\it Gaia} DR2 software pipeline are 
        described in \citet{2019A&A...627A.104D}. 

     \subsection{Numerical integrations}
        In order to study the orbital evolution of C/2018~V1 and other previously detected minor bodies following hyperbolic paths, we have 
        used a direct $N$-body code originally written by S.~J.~Aarseth that implements a fourth-order version of the Hermite integration 
        scheme \citep{1991ApJ...369..200M,2003gnbs.book.....A}. The standard version of this code is publicly available from the Institute 
        of Astronomy web site.\footnote{\url{http://www.ast.cam.ac.uk/~sverre/web/pages/nbody.htm}} The numerical integrations of the orbit 
        of C/2018~V1 consider the perturbations by eight major planets and treat the Earth--Moon system as two separate objects; they also 
        include the barycentre of the dwarf planet Pluto--Charon system and the three most massive asteroids of the main belt, namely, dwarf 
        planet (1) Ceres, (2) Pallas, and (4) Vesta. Additional details of the code used in our research and of our integrations and 
        physical model are described in \citet{2012MNRAS.427..728D} and \citet{2018MNRAS.476L...1D}. Our calculations do not include 
        non-gravitational forces. The orbit determinations in Table~\ref{elements} did not require non-gravitational terms to fit the 
        available astrometry; this suggests that any contribution due to asymmetric outgassing is probably a second order effect in this 
        case. Therefore, we believe that neglecting the role of non-gravitational forces in our integrations is unlikely to have any major 
        impact on our conclusions. The uncertainties in the values of the orbital elements, see Table~\ref{elements}, have been included in 
        our simulations by applying the covariance matrix methodology discussed by \citet{2015MNRAS.453.1288D}. Calculations of heliocentric 
        Galactic space velocities have been carried out as described by \citet{1987AJ.....93..864J} using the values of the relevant 
        parameters provided by \citet{2010MNRAS.403.1829S}, and applying the Monte Carlo sampling procedure described by 
        \citet{2019A&A...627A.104D}. Averages, standard deviations, and other statistics have been computed in the usual way (see e.g. 
        \citealt{2012psa..book.....W}).

     \subsection{A matter of orbital uncertainty?}
        As pointed out above, most minor bodies following putative nearly hyperbolic paths never attract enough attention to get good orbit 
        determinations. In addition, it is customary that whenever discussing the possibility that an object has a hyperbolic orbit one 
        should refer to the eccentricity of the barycentric orbital elements; most comets with values of the heliocentric eccentricity 
        $\geq1$ have barycentric values $<1$. For example, the orbit determination of C/2008~J4 (McNaught) is based on a data-arc span of 
        15~d with a heliocentric eccentricity of 1.008$\pm$0.010 but a barycentric value of 1.001 and C/2012~S1 (ISON) has a data-arc span
        of 784~d with a heliocentric value of 1.00020$\pm$0.00002 and a barycentric one of 0.99957 (see Table~\ref{candi}). However, all 
        these values are osculating ones, often corresponding to the midpoint epoch of the available data-arc. These values may change over 
        time as the objects interact with massive bodies in the Solar system such as the Sun or Jupiter. Besides C/2018~V1 (see 
        Table~\ref{elements}), another rare example of hyperbolic comet with both hyperbolic heliocentric and barycentric orbit 
        determinations is C/1997~P2 (Spacewatch) with a data-arc span of 49~d, a heliocentric eccentricity of 1.0279$\pm$0.0002, and a 
        barycentric one of 1.0182 ---i.e. it is hyperbolic at the 133$\sigma$ level heliocentrically and at the 87$\sigma$ level 
        barycentrically (see Table~\ref{candi}). 
%
%
     \begin{table*}
        \centering
         \fontsize{8}{11pt}\selectfont
         \tabcolsep 0.25truecm
         \caption{\label{candi} Candidate insterstellar comets and 1I/2017~U1 (`Oumuamua) as discussed by \citet{2018MNRAS.476L...1D}. Data
                  include the object's designation, the heliocentric eccentricity and its 1$\sigma$ uncertainty, 
                  $e_{\rm h}\pm\sigma_{e_{\rm h}}$, the barycentric eccentricity, $e_{\rm b}$, the statistical significance of $e_{\rm b}$,
                  ($e_{\rm b}$ - 1)/$\sigma_{e_{\rm h}}$ that has been computed using all the decimal figures provided by the data source, 
                  the number of observations, and the data-arc span. The values of the eccentricity are referred to the epoch used by JPL's 
                  SBDB, which is different for each object. Source: JPL's SBDB. 
                 }
         \begin{tabular}{lccccc}
          \hline\hline
            Object                                & $e_{\rm h}$           & $e_{\rm b}$ &  sig.    &  obs.   &  span \\
                                                  &                       &             &          &         &  (d)  \\
          \hline
            C/1853 R1 (Bruhns)                    & 1.000664$\pm$n/a      &  1.002029   &   n/a    &   17    &   90  \\ 
            C/1997~P2 (Spacewatch)                & 1.0279$\pm$0.0002     &  1.0182     & 86.98    &   94    &   49  \\
            C/1999~U2 (SOHO)                      & 1.0$\pm$0.3           &  0.88       & $-$0.39  &   41    &    1  \\
            C/2002~A3 (LINEAR)                    & 1.00949$\pm$0.00002   &  1.00290    & 135.73   &  285    &  527  \\
            C/2008~J4 (McNaught)                  & 1.008$\pm$0.010       &  1.001      & 0.11     &   22    &   15  \\
            C/2012~C2 (Bruenjes)                  & 1.0034$\pm$0.0009     &  1.0001     & 0.13     &  246    &   16  \\
            C/2012~S1 (ISON)                      & 1.00020$\pm$0.00002   &  0.99957    & $-$21.50 & 6514    &  784  \\ 
            C/2017~D3 (ATLAS)                     & 1.002434$\pm$0.000009 &  1.000301   & 32.07    &  520    &  773  \\
            C/2018~V1 (Machholz-Fujikawa-Iwamoto) & 1.00040$\pm$0.00002   &  1.01235    & 510.93   &  750    &   37  \\ 
            1I/2017~U1 (`Oumuamua)                & 1.20113$\pm$0.00002   &  1.20315    & 9673.71  &  207    &   80  \\  
          \hline
         \end{tabular}
     \end{table*}
%
%

        Statistically, the bound orbit ($e<1$) is the null hypothesis and solid evidence needs to be provided if the claim is that the orbit 
        is in fact hyperbolic. Osculating hyperbolic orbit determinations (heliocentric and barycentric) with a sufficiently high $\sigma$ 
        level may be indicative of an extrasolar origin, but to actually show that their provenance is outside the Solar system $N$-body 
        calculations must be used. In summary and in order to validate such theoretical expectations ---i.e. that the null hypothesis can be 
        rejected--- a representative set of control orbits (statistically compatible with the available observations) must be integrated 
        forward and backwards in time to confirm that the dynamical evolution of the candidate over a reasonable amount of time (a few 
        hundred thousand years for relatively slow objects, tens of thousands for those as fast or faster than `Oumuamua) is consistent with 
        not being bound to the Solar system ---i.e. their relative velocity is above 0.5~km~s$^{-1}$ when near the Hill radius of the Solar 
        system (see e.g. \citealt{1982ApJ...255..307V}). This methodology has been previously used by \citet{2018RNAAS...2b..10D} to confirm 
        independently that C/2017~K2 (PANSTARRS) is a bound and dynamically old Oort Cloud comet ---see the works by 
        \citet{2018AJ....155...25H} and \citet{2018A&A...615A.170K}--- as well as to show that C/2018~F4 (PANSTARRS) could be a genuine 
        representative of the average Oort Cloud comet population \citep{2019A&A...625A.133L}.
        
  \section{Comet C/2018~V1 (Machholz-Fujikawa-Iwamoto): the current orbit in context}
\begin{figure}
        \centering
        \includegraphics[width=\linewidth]{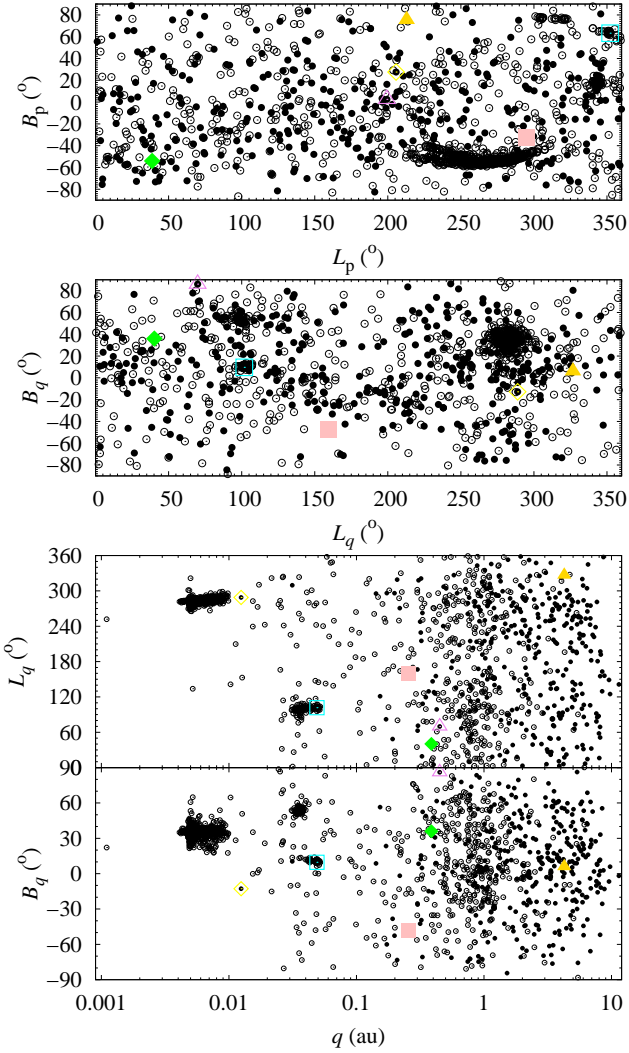}
        \caption{Poles (top panel) and perihelia (middle and bottom panels) of known parabolic (black empty circles) and hyperbolic (black 
                 filled circles) minor bodies (2191 objects); C/2018~V1 (Machholz-Fujikawa-Iwamoto) is plotted as a green filled diamond. 
                 The large cluster centred at $(L_{\rm q}, B_{\rm q})\sim(282\degr, 35\degr)$ and $(L_{\rm p}, B_{\rm p})\sim(269\degr, 
                 -51\degr)$ signals the Kreutz family of comets (with $q<0.01$~au); the other clusters at $L_{\rm q}\sim100\degr$ are 
                 associated with the various groups of SOHO comets (Meyer, Marsden and Kratch). The objects plotted in colour are 1I/2017~U1 
                 (`Oumuamua) in pink (filled square), C/1997~P2 (Spacewatch) in amber (filled triangle), C/1999~U2 (SOHO) in cyan (empty 
                 square), C/2008~J4 (McNaught) in violet (empty triangle), and C/2012~S1 (ISON) in yellow (empty diamond).
                }
        \label{pps}
     \end{figure}

     The overall orientations in space of the orbits of parabolic and hyperbolic minor bodies can be studied using the coordinates of their 
     perihelia and poles. In heliocentric ecliptic coordinates, the longitude and latitude of the orbital pole are $(L_{\rm p}, B_{\rm p}) = 
     (\Omega-90\degr, 90\degr-i)$. The ecliptic coordinates of the perihelion, $(L_q, B_q)$, are given by the expressions: 
     $\tan{(L_q-\Omega)}=\tan\omega\,\cos{i}$ and $\sin{B_q}=\sin\omega\,\sin{i}$ (see e.g. \citealt{1999ssd..book.....M}). Figure~\ref{pps} 
     shows the distributions in the sky of the poles (top panel) and perihelia (middle panel) as well as the perihelion distances (bottom 
     panel) of the known comets moving in parabolic (black empty circles) or hyperbolic (black filled circles) orbits (2191 objects as of 
     2019 July 26, 354 with $e>1$). The values of $i$, $\Omega$ and $\omega$ are practically independent of the nature, heliocentric or
     barycentric, of the orbit determination (see e.g. \citealt{2017MNRAS.471L..61D}). 

     Some clusters are clearly visible in Fig.~\ref{pps} and they are all associated with sungrazing comet groups \citep{1967AJ.....72.1170M,
     1989AJ.....98.2306M}; they include the Kreutz group comets \citep{1888uudc.book.....K} and the Meyer, Marsden and Kracht (1 and 2) 
     group comets \citep{2002ApJ...566..577S,2005ARA&A..43...75M,2008PhDT........14K,2013ApJ...778...24S,2015ApJ...815...52S,
     2016ApJ...823....2S}. Kreutz sungrazers are found at $(L_q, B_q)\sim(282\degr, 35\degr)$ and they could be the result of cascading 
     fragmentation taking place during the last 1700 years \citep{2004ApJ...607..620S,2007ApJ...663..657S}, Meyer group comets have 
     $(L_q, B_q)\sim(98\degr, 53\degr)$, Marsden group comets have $(L_q, B_q)\sim(102\degr, 10\degr)$, and Kracht (1 and 2) group comets 
     have $(L_q, B_q)\sim(102\degr, 11\degr)$ ---Marsden comets have $i\sim26\degr$, but Kracht comets have $i\sim13\degr$. The distribution 
     in $q$ (bottom panels) is clearly different for parabolic and hyperbolic comets: parabolic comets tend to have shorter perihelia if we 
     exclude those of the Kreutz family ---Fig.~\ref{pps}, bottom panel, shows that the perihelion distances of Kreutz sungrazers are 
     $<0.01$~au, but $>0.004$~au. This is to be expected if parabolic comets have experienced multiple perihelion passages. 
%
%
%
%

     Figure \ref{pps} shows that comet C/2018~V1 (Machholz-Fujikawa-Iwamoto) ---plotted as a green filled diamond, $(L_{\rm p}, B_{\rm p}) = 
     (38\fdg72, -53\fdg99)$ and $(L_q, B_q) = (40\fdg24, 36\fdg00)$--- does not belong to any of the sungrazing comet groups. In fact, it 
     does not appear to be dynamically coherent with any of the known parabolic or hyperbolic comets, which might indicate that it is not a 
     first time visitor from the Oort Cloud \citep{2018MNRAS.476L...1D}. Furthermore, the probability of finding another minor body within 
     15{\degr} of C/2018~V1 in both the $(L_{\rm p}, B_{\rm p})$ and $(L_q, B_q)$ planes is 0.00137 (3/2190). The closest comets in terms of 
     orbital geometry are C/1857~V1 (Donati-van Arsdale), C/1948~T1 (Wirtanen), and C/1997~S4 (SOHO), but they are several degrees away in 
     terms of poles and perihelia; within 11{\degr} there are none and within 13{\degr} there are 2 comets, C/1857~V1 and C/1997~S4. Figure 
     \ref{pps} also shows the location of other relevant objects to be discussed later; one of them, C/1999~U2 (SOHO) that is plotted as an
     cyan empty square, could be a {\it bona fide} Marsden comet as $(L_q, B_q)=(101.8\degr, 9.6\degr)$ and $i=26.8\degr$, not a candidate 
     interstellar comet as suggested by \citet{2018MNRAS.476L...1D}.

   \section{Comet C/2018~V1 (Machholz-Fujikawa-Iwamoto): dynamical evolution}
\begin{figure}
         \centering
         \includegraphics[width=\linewidth]{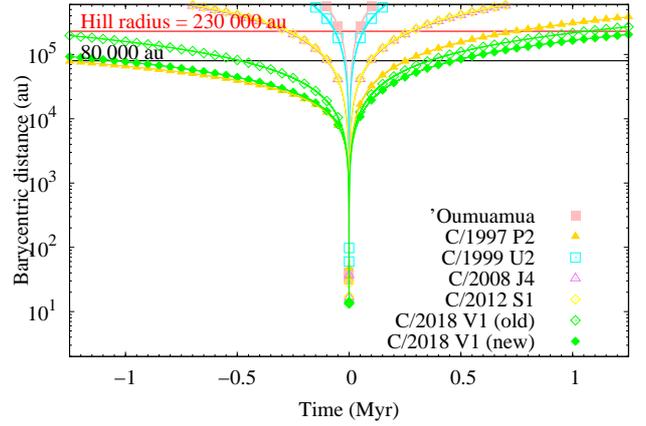}
         \caption{Evolution of the barycentric distance of 1I/2017~U1 (`Oumuamua), plotted in pink (filled squares), C/1997~P2 (Spacewatch) 
                  in amber (filled triangles), C/1999~U2 (SOHO) in cyan (empty squares), C/2008~J4 (McNaught) in violet (empty triangles), 
                  C/2012~S1 (ISON) in yellow (empty diamonds), and C/2018~V1 (Machholz-Fujikawa-Iwamoto) in green (old solution, empty 
                  diamonds; new solution, filled diamonds) ---all based on nominal orbit determinations; the zero instant of time 
                  corresponds to epoch JDTDB~2458600.5, 27-April-2019. The evolutions of C/2008~J4 and C/2012~S1 closely overlap.
                 }
         \label{evolution}
      \end{figure}

      Figure~\ref{evolution} shows the (past and future) short-term orbital evolution of comet C/2018~V1 (Machholz-Fujikawa-Iwamoto) ---in 
      green filled diamonds, new nominal orbit in Table~\ref{elements}--- and those of a few hyperbolic minor bodies of probable or possible 
      interstellar origin \citep{2018MNRAS.476L...1D} ---namely, 1I/2017~U1 (`Oumuamua) in pink filled squares (nominal heliocentric 
      $e=1.20113\pm0.00002$, 207 observations, data-arc span 80~d), C/1997~P2 (Spacewatch) in amber filled triangles ($e=1.0279\pm0.0002$, 
      94 observations, data-arc span 49~d), C/1999~U2 (SOHO) in cyan empty squares ($e=1.0\pm0.3$, 41 observations, data-arc span 1~d), 
      C/2008~J4 (McNaught) in violet empty triangles ($e=1.008\pm0.010$, 22 observations, data-arc span 15~d), and C/2012~S1 (ISON) in 
      yellow empty diamonds ($e=1.00020\pm0.00002$, 6514 observations, data-arc span 784~d), but these last two nearly overlap. The black
      thick line corresponds to the aphelion distance ---$a \ (1 + e)$, limiting case $e=1$--- that defines the domain of dynamically old 
      Oort Cloud comets (i.e. semimajor axis $a < 40 000$~au, see \citealt{2017MNRAS.472.4634K}) as opposed to those that may be dynamically 
      new, or first time visitors from the Oort Cloud; the red thick line signals the radius of the Hill sphere of the Solar system (see 
      e.g. \citealt{1965SvA.....8..787C}). Figure~\ref{evolution} shows that, besides C/2018~V1, C/1997~P2, C/2008~J4 and C/2012~S1 are 
      suitable candidates to be interstellar comets (see Table~\ref{candi} for additional details); as pointed out above, C/1999~U2 is a 
      probable Marsden group comet and its orbit determination is perhaps too uncertain to make it a candidate, even if it is second to 
      `Oumuamua in terms of dynamical evolution. For comparison, results from the old nominal orbit of C/2018~V1 in Table~\ref{elements} are 
      plotted as green empty diamonds.
%
%
%
%

\begin{figure*}
        \centering
         \includegraphics[width=0.49\linewidth]{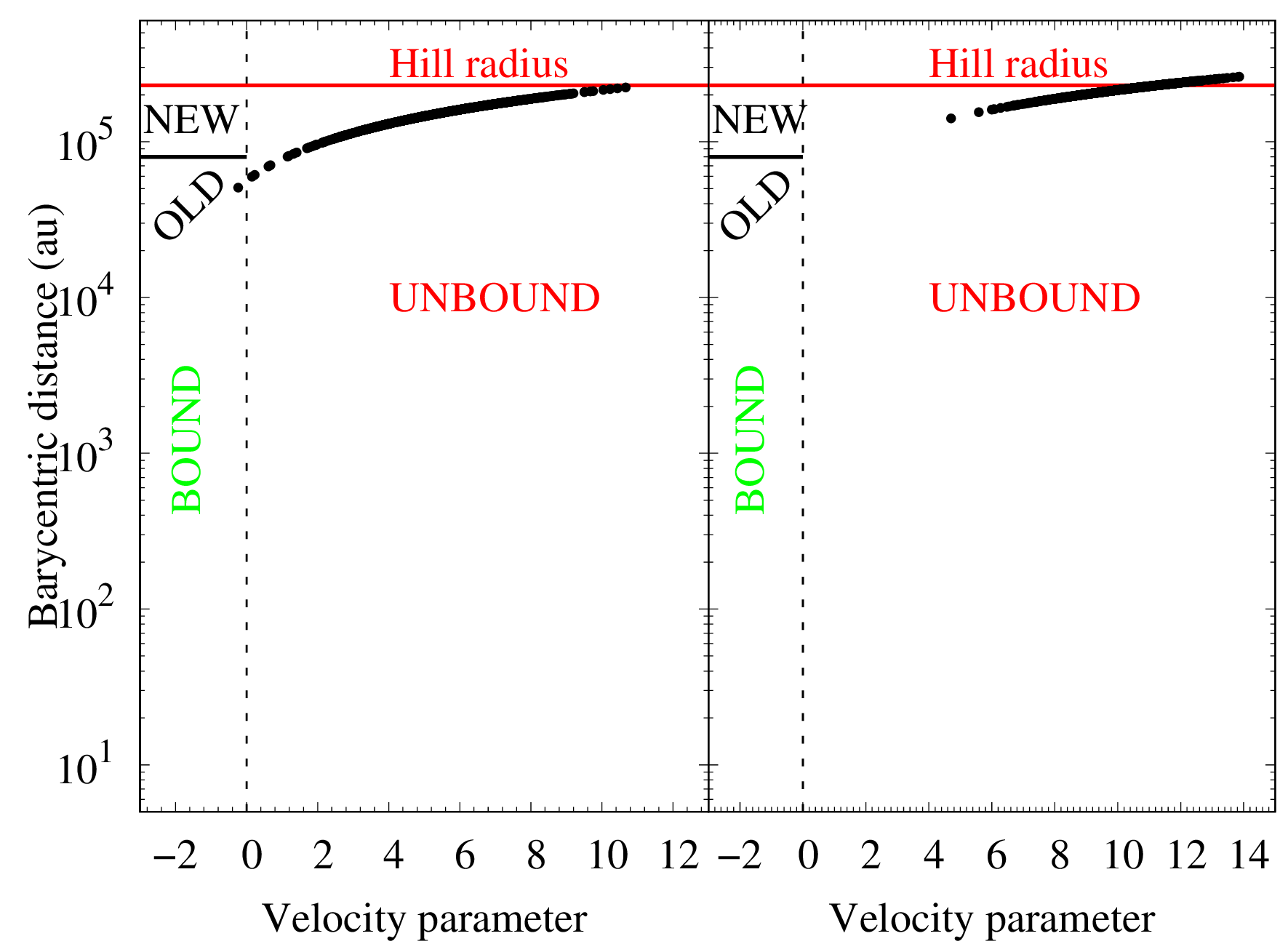}
         \includegraphics[width=0.49\linewidth]{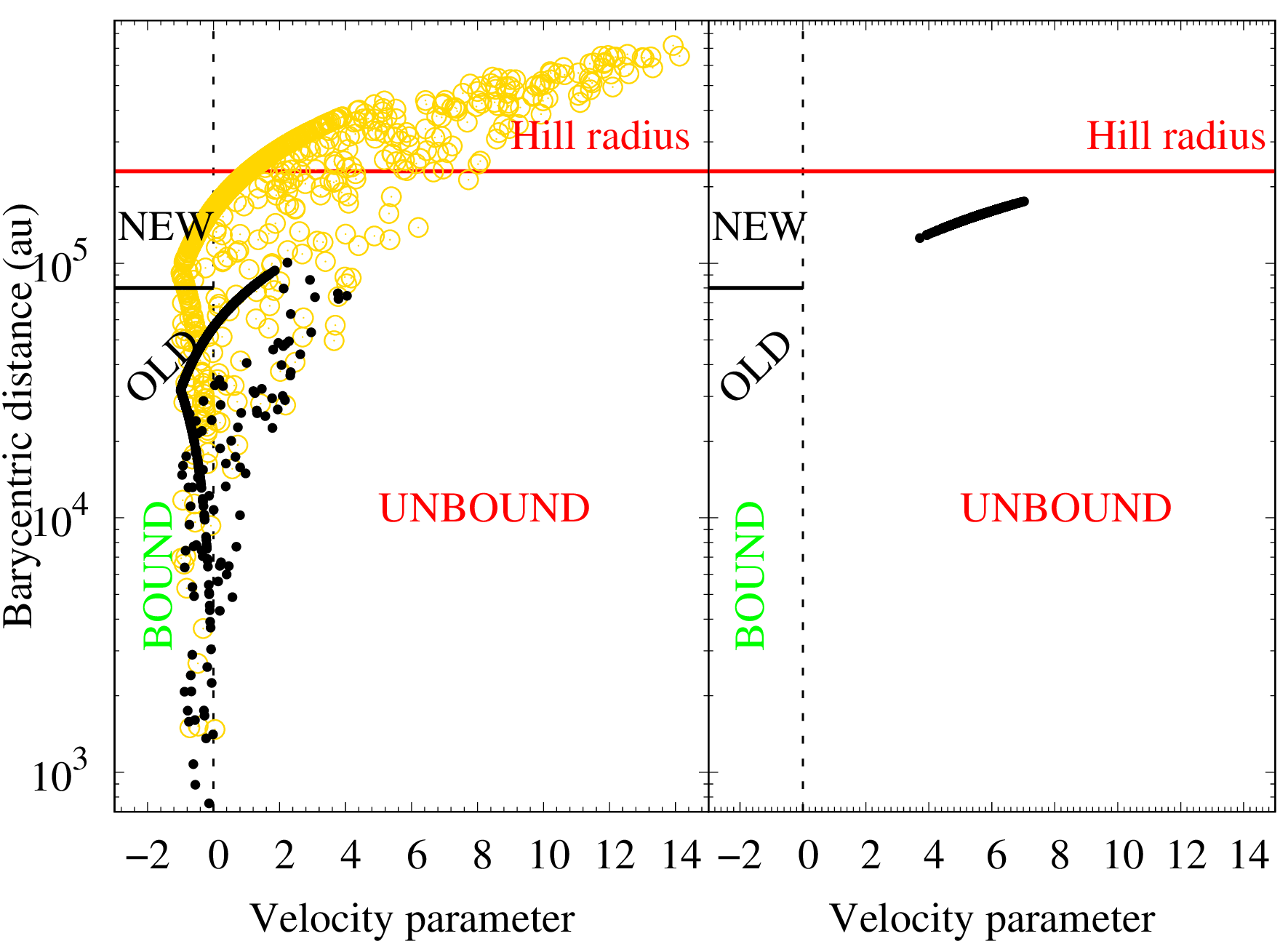}
         \caption{Values of the barycentric distance as a function of the velocity parameter. The left-hand side set of two panels 
                  corresponds to results from the first orbit determination (2018 November 20) shown in Table~\ref{elements}; the right-hand 
                  side set shows results from the new orbit determination (2019 May 15). In both cases, the panels show the outcome of the 
                  evolution 1~Myr into the past (left-hand side panel) and 1~Myr into the future (right-hand side panel) for 1000 control 
                  orbits of C/2018~V1 (Machholz-Fujikawa-Iwamoto), black filled circles. For the new orbit, the results of 700 control orbits 
                  evolved 5~Myr into the past, amber empty circles, are also shown.
                 }
         \label{inout}
      \end{figure*}

      Figure~\ref{inout} shows the barycentric distance as a function of the velocity parameter for 1000 control orbits of C/2018~V1. The 
      left-hand side set of panels shows results from integrations with input parameters consistent with the old solution in 
      Table~\ref{elements} (2018 November 20); the ones on the right-hand side show results from input data based on the new orbit 
      determination (2019 May 15). In both cases, the left-hand side panel shows results at 1~Myr into the past and the right-hand side 
      panel displays those at 1~Myr into the future. The velocity parameter is the difference between the barycentric and escape velocities 
      at the computed barycentric distance in units of the escape velocity. Positive values of the velocity parameter identify control 
      orbits that could have been followed by objects of interstellar provenance (left-hand side panel) or that lead to ejections from the 
      Solar system (right-hand side panel). The initial conditions of the control orbits have been generated using the covariance matrix as 
      described above. 

      For the old orbit determination (Fig.~\ref{inout}, left-hand side set of panels), the vast majority, $>99.9$ per cent, of the control 
      orbits computed here placed C/2018~V1 close to or beyond the Hill sphere 1.3~Myr ago and 1.3~Myr into the future as well, i.e. under 
      the gravitational influence of the Galactic tide. In summary and considering the old orbit determination in Table~\ref{elements}, our 
      $N$-body simulations and statistical analyses suggest that C/2018~V1 came from interstellar space and it will return back to it, 
      C/2018~V1 cannot be a dynamically new or old Oort Cloud comet (compare Figs~\ref{evolution} and \ref{inout} with fig.~5 in 
      \citealt{2019A&A...625A.133L}).

      When input data based on the new orbital solution shown in Table~\ref{elements} are used to compute the past and future orbital 
      evolution of C/2018~V1 (Fig.~\ref{inout}, right-hand side set of panels, black points), about 43 per cent of the 1~Myr integrations 
      into the past are compatible with an origin in interstellar space, and all the experiments performed predict that C/2018~V1 will be 
      unbound from the Solar system, 1~Myr into the future. When longer integrations are carried out (Fig.~\ref{inout}, right-hand side set 
      of panels, amber empty circles, 5~Myr into the past), the probability of having an extrasolar origin increases to about 73 per cent
      (72.6$\pm$0.5). This variation strongly suggests that this object, if extrasolar, may have remained weakly bound to the Solar system 
      for a certain amount of time before returning now to interstellar space.

      With these results in mind, C/2018~V1 may be a former member of the Oort Cloud that was perturbed a few million of years ago into the 
      dynamically unstable path that now follows, but an origin beyond the Solar system cannot be excluded at any significant level by our 
      analysis considering the orbit determination currently available for this object (see Table~\ref{elements}). However, if C/2018~V1 has 
      an extrasolar origin, it arrived in the Solar system with a relative velocity barely above the minimum one to become unbound. With 
      these facts laid on the table, in the following section we explore a plausible what-if scenario. This hypothetical scenario aims to 
      respond to the question, can a very slow interstellar comet candidate still be compatible with an origin beyond the Oort Cloud?  
%
%
%
%

   \section{Comet C/2018~V1 (Machholz-Fujikawa-Iwamoto): an origin among the solar siblings?}
      The kinematic properties of minor bodies escaped from a planetary system hosted by another star must be consistent with those of stars 
      in the solar neighbourhood; therefore, investigating the pre-perihelion trajectory of comet C/2018~V1 (Machholz-Fujikawa-Iwamoto) may 
      shed some light on its true origin. The purpose of this section is not to single out the possible extrasolar source (i.e. the actual 
      star) of C/2018~V1, but to investigate the plausibility of an origin among the known stars in the solar neighbourhood. In other words, 
      if there are no known stars with a kinematic signature compatible with that of C/2018~V1, when the control orbit leads to an unbound 
      state 1~Myr into the past, then the former Oort Cloud membership automatically becomes the preferred scenario for the origin of this 
      comet. However, if there are stars that can be considered as kinematic analogues of C/2018~V1 in its unbound version, then an 
      extrasolar origin cannot be fully excluded.

      A statistical analysis of the results of the simulations shown in Fig.~\ref{inout} indicates that at 0.31$\pm$0.08~pc from the 
      barycentre of the Solar system and 1~Myr into the past, C/2018~V1 was moving inwards, at $-$0.30$\pm$0.14~km~s$^{-1}$ ---i.e. 
      below the 0.5~km~s$^{-1}$ critical value pointed out by \citet{1982ApJ...255..307V}--- and projected towards 
      $\alpha=13^{\rm h}~24^{\rm m}~36^{\rm s}$, $\delta=-48\degr~03\arcmin~00\arcsec$ $(201\fdg2\pm0\fdg3, -48\fdg05\pm0\fdg13)$ in the 
      constellation of Centaurus (geocentric radiant or antapex) with Galactic coordinates $l=308\fdg64$, $b=+14\fdg43$, and ecliptic 
      coordinates $\lambda=219\fdg69$, $\beta=-35\fdg92$. The components of its heliocentric Galactic velocity were 
      $(U, V, W)=(-0.18\pm0.08, 0.22\pm0.10, -0.08\pm0.04)$~km~s$^{-1}$, which are compatible with an origin in a star with very small, 
      but not zero, relative motion with respect to the Sun. If the 5~Myr-into-the-past calculations are used, the results are consistent 
      with the previous ones: C/2018~V1 was located at 1.3$\pm$0.6~pc from the barycentre of the Solar system and moving inwards, at 
      $-$0.4$\pm$0.2~km~s$^{-1}$, projected towards $\alpha=13\fh3\pm0\fh7$ and $\delta=-48\degr\pm2\degr$.

      On the other hand, at 0.74$\pm$0.04~pc from the Sun and 1~Myr into the future, C/2018~V1 will be receding from us at 
      0.70$\pm$0.04~km~s$^{-1}$ ---this time exceeding the 0.5~km~s$^{-1}$ critical value pointed out by \citet{1982ApJ...255..307V}--- 
      towards $\alpha=13^{\rm h}~33^{\rm m}~58^{\rm s}$, $\delta=-48\degr~45\arcmin~36\arcsec$ ($203\fdg49\pm0\fdg09, -48\fdg76\pm0\fdg03)$ 
      also in the constellation of Centaurus (apex) with Galactic coordinates $l=310\fdg12$, $b=+13\fdg52$, and ecliptic coordinates 
      $\lambda=221\fdg78$, $\beta=-35\fdg83$. Its post-perihelion heliocentric Galactic velocity will be (0.44$\pm$0.02, $-$0.52$\pm$0.03, 
      0.164$\pm$0.009)~km~s$^{-1}$.

\begin{figure}
        \centering
         \includegraphics[width=\linewidth]{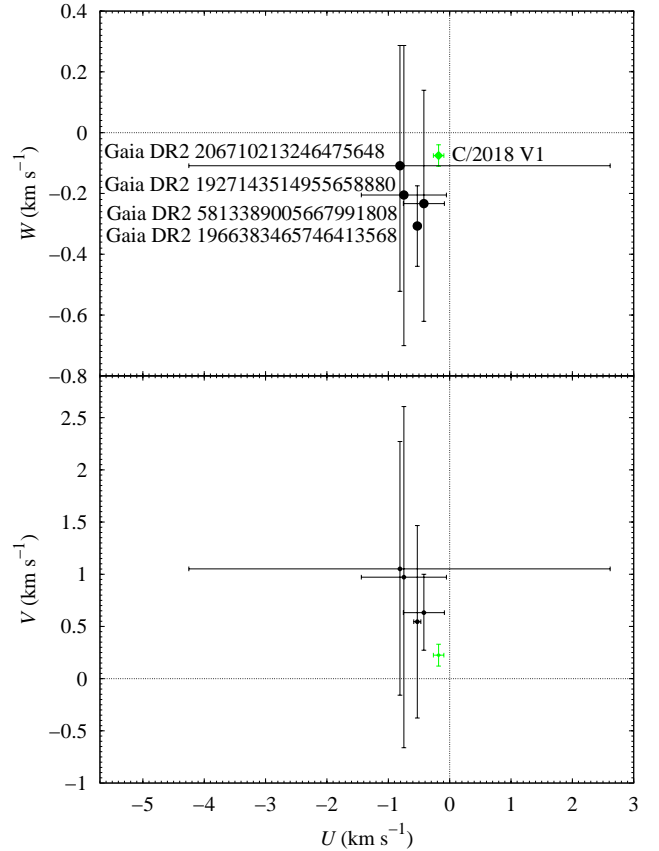}
         \caption{Heliocentric Galactic velocity components of C/2018~V1 (Machholz-Fujikawa-Iwamoto), plotted in green (filled diamond), and 
                  four stars with values of their velocity components consistent within 9$\sigma$ with those of the comet (see 
                  Table~\ref{stars}). The stellar input data used to prepare this figure are from {\it Gaia} DR2.
                 }
         \label{UVW}
      \end{figure}

      Considering the sample described in Section~2.2 and looking for stars with relative errors in the value of the parallax better than 20 
      per cent, we have found four with values of their heliocentric Galactic velocity components consistent ---within 9$\sigma$--- with 
      those of the comet when inbound; their properties are shown in Tables~\ref{stars} and \ref{starsuvw}---distances in 
      Table~\ref{starsuvw} are from \citet{2018AJ....156...58B}--- and their kinematic signatures are plotted together with that of the 
      comet in Fig.~\ref{UVW}. Only one of them is from the solar neighbourhood (within 100~pc from the Sun). The stars in Tables~\ref{stars} 
      and \ref{starsuvw} are relatively unstudied. A search for matching sources has been carried out using the tools provided by 
      VizieR\footnote{\url{http://vizier.u-strasbg.fr/viz-bin/VizieR}} \citep{2000A&AS..143...23O} with a radius of 1\farcs0. 

      Given the what-if nature of the analysis performed in this section, only the most relevant additional data are discussed here. 
      {\it Gaia} DR2 206710213246475648 has been observed by the Large Sky Area Multi-Object Fibre Spectroscopic Telescope (LAMOST) 
      spectroscopic surveys \citep{2012RAA....12.1197C,2012RAA....12..723Z}\footnote{\url{http://dr4.lamost.org}} as target 
      J050221.30+455624.0; relevant data include a value of the heliocentric radial velocity of 0.23$\pm$4.67~km~s$^{-1}$, a metallicity 
      [Fe/H] of $-$0.005$\pm$0.020, and a stellar sub-class of G3 (the Sun has a spectral type/luminosity class of G2V). {\it Gaia} DR2 
      1927143514955658880 has also been observed by LAMOST, its target identification is J235459.91+461605.9; its heliocentric radial 
      velocity is given as 0.25$\pm$5.67~km~s$^{-1}$, its metallicity is $-$0.113$\pm$0.066, and its stellar sub-class is G3. {\it Gaia} DR2 
      1966383465746413568 is also known as TYC 3191-276-1 and {\it Gaia} DR2 5813389005667991808 is TYC 9064-2770-1; both stars are included 
      in the Tycho-2 catalogue \citep{2000A&A...355L..27H}. Although it would have been desirable to have less uncertain values of the 
      radial velocities of our sources, our search in VizieR suggests that these four sources remain poorly studied and the few available 
      data are of inferior quality when compared with those available from {\it Gaia} DR2 (see values of the radial velocities in 
      Table~\ref{stars}). This is to be expected, the stars of interest here have very low relative velocities with respect to the Sun; 
      consistently, the values of their radial velocity are low and their relative errors large (see Table~\ref{stars}). 
      
      The fact is that if C/2018~V1 has an extrasolar provenance, it may have approached the Solar system at low relative velocity with 
      respect to the Sun and this places an implicit connection between the interplanetary and interstellar environments. The Sun was born 
      within a star cluster (see e.g. \citealt{2010ARA&A..48...47A}). It is still under debate whether this cluster was gravitationally 
      bound ---open cluster (see e.g. \citealt{2009ApJ...696L..13P})--- or unbound ---a stellar association (see e.g. 
      \citealt{2013A&A...549A..82P}). Both stellar associations and open clusters eventually dissolve, contributing to the field stellar 
      populations (see e.g. \citealt{2008ApJ...672..342D}). The search for solar siblings (see e.g. \citealt{2016PhDT.......183M}) or stars 
      that formed together with the Sun has made steady progress during the last decade or so. In order to be classified as a solar sibling 
      candidate, a nearby star must have age and chemical abundances (metallicity and isotopic ratios) consistent with those of the Sun (see 
      e.g. \citealt{2018A&A...619A.130A}). 

      In principle, a putative solar sibling should also have a small space motion relative to the Sun; however, if the Sun ---together with 
      many other physically unrelated stars--- is trapped in a spiral corotation resonance \citep{2017ApJ...843...48L}, having similar 
      kinematics is no longer a robust condition to qualify as a solar sibling. \citet{2009ApJ...696L..13P} has pointed out that a small 
      number of solar siblings may still remain in the neighbourhood of the Sun, which triggered searches for suitable candidates (see e.g. 
      \citealt{2015A&A...575A..51L,2016MNRAS.457.1062M}). However, \citet{2011MNRAS.412.1771M} argued that this is 
      unlikely when considering the long sequence of secular perturbations experienced by these stars in their journey throughout the 
      Galactic disc. On the other hand, \citet{2015CeMDA.121..107V} pointed out that less than 10 percent ---and probably just about 1 per 
      cent--- of the true solar siblings could still remain within 100~pc of the present position of the Sun.
      In any case, stars with small space motions relative to the Sun, be they solar siblings or not, may host structures 
      similar to our Oort Cloud \citep{1950BAN....11...91O} that may leak comets into the interstellar medium. Such minor bodies may 
      experience hyperbolic encounters with the Solar system, entering from interstellar space, and be eventually detected from the Earth as 
      slightly hyperbolic comets. None of the stars in Table~\ref{stars} are listed as solar sibling candidates \citep{2018A&A...619A.130A}. 
      It is unclear whether C/2018~V1 might have had an origin in any of them (if we assume that it has an extrasolar provenance, which may 
      be the most likely interpretation, statistically), but they are reasonably good kinematic analogues of C/2018~V1. Having been able to
      find several relatively good kinematic analogues of C/2018~V1 among those stars relatively close to the Sun only means that the 
      predicted kinematic signature of C/2018~V1 prior to its recent perihelion passage (if originally unbound) is consistent with that of 
      observed stars, i.e. it is not unphysical. On the other hand, the Solar system departure kinematics of C/2018~V1 shows that minor 
      bodies can leave the sphere of influence of a planetary system at a very low relative speed, which may eventually become the approach 
      velocity when the same object experiences a close encounter with another planetary system. Within this context, C/2018~V1 could be the 
      first example of a new class of comets discussed by \citet{2019arXiv190610617T}, the transitional interstellar comets. 
%
%
     \begin{table*}
        \centering
         \fontsize{8}{11pt}\selectfont
         \tabcolsep 0.10truecm
         \caption{\label{stars}Kinematic matches of C/2018~V1 (Machholz-Fujikawa-Iwamoto) from {\it Gaia} DR2 (I). {\it Gaia} DR2 
                  designation, $\alpha$, $\delta$, $\pi$, $\sigma _{\pi}$, $\mu _{\alpha}$, $\sigma _{\mu_{\alpha}}$, $\mu _{\delta}$, 
                  $\sigma _{\mu_{\delta}}$, $V_{\rm r}$, and $\sigma_{V_{\rm r}}$ from {\it Gaia} DR2.  
                 }
         \begin{tabular}{rcccccccccc}
          \hline\hline
            {\it Gaia} DR2 designation & $\alpha$        & $\delta$          & $\pi$   & $\sigma_{\pi}$ & $\mu_{\alpha}$  & $\sigma_{\mu_{\alpha}}$ &
                                         $\mu_{\delta}$  & $\sigma_{\mu_{\delta}}$ & $V_{\rm r}$   & $\sigma_{V_{\rm r}}$ \\
                                       &  (\degr)        &  (\degr)          & (mas)   & (mas)          & (mas~yr$^{-1}$) & (mas~yr$^{-1}$)         &
                                         (mas~yr$^{-1}$) & (mas~yr$^{-1}$)         & (km~s$^{-1}$)  & (km~s$^{-1}$)       \\
          \hline
            206710213246475648         &  75.58874228476 &   +45.93995550159 & 15.8998 & 0.8093         & $-$1.921        & 1.275                   &
                                          1.612          & 1.111                   &    1.13        &  3.66               \\
           1927143514955658880         & 358.74965337143 &   +46.26832670029 & 2.7573  & 0.0290         &    0.159        & 0.041                   &
                                          0.107          & 0.030                   &    1.21        &  1.86               \\
           1966383465746413568         & 323.37898817647 &   +41.72653005266 & 3.2308  & 0.0236         &    0.382        & 0.037                   &
                                          0.135          & 0.037                   &    0.56        &  0.93               \\
           5813389005667991808         & 259.80441797335 & $-$66.38996438375 & 5.3887  & 0.2645         &    0.578        & 0.387                   &
                                          0.026          & 0.516                   & $-$0.61        &  0.25               \\
          \hline
         \end{tabular}
     \end{table*}
%
%
%
%
     \begin{table}
        \centering
         \fontsize{8}{11pt}\selectfont
         \tabcolsep 0.10truecm
         \caption{\label{starsuvw}Kinematic matches of C/2018~V1 (Machholz-Fujikawa-Iwamoto) from {\it Gaia} DR2 (II). {\it Gaia} DR2 
                  designation, $d$ from \citet{2018AJ....156...58B}, heliocentric Galactic velocity components ($U$, $V$, $W$) computed as 
                  described in the text.
                 }
         \begin{tabular}{rcccc}
          \hline\hline
            {\it Gaia} DR2 designation & $d$              & $U$              & $V$           & $W$              \\
                                       & (pc)             & (km~s$^{-1}$)    & (km~s$^{-1}$) & (km~s$^{-1}$)    \\
          \hline
            206710213246475648         & 63$^{+4}_{-3}$  & $-$0.84$\pm$3.46 & 1.06$\pm$1.22 & $-$0.11$\pm$0.40 \\
           1927143514955658880         & 359$^{+4}_{-4}$  & $-$0.75$\pm$0.70 & 0.98$\pm$1.65 & $-$0.21$\pm$0.50 \\
           1966383465746413568         & 307$^{+2}_{-2}$  & $-$0.53$\pm$0.06 & 0.54$\pm$0.92 & $-$0.31$\pm$0.13 \\
           5813389005667991808         & 185$^{+10}_{-8}$ & $-$0.42$\pm$0.34 & 0.63$\pm$0.37 & $-$0.24$\pm$0.37 \\
          \hline
         \end{tabular}
     \end{table}
%
%
%
%
%
%

  \section{Discussion}
      The current heliocentric and barycentric orbit determinations of C/2018~V1 (Machholz-Fujikawa-Iwamoto) in Table~\ref{elements} can be 
      considered as compatible with this comet being unbound from the Solar system. However, considering all the available data and their 
      associated uncertainties we have to conclude that they all appear to point in the same direction: rather than having come from 
      interstellar space, C/2018~V1 seems to have been dislodged from the Oort Cloud in the recent past (a few Myr ago). Figure \ref{pps} 
      can be interpreted as an indication that C/2018~V1 is part of the essentially isotropic Oort cloud background population, but the fact 
      remains that material from interstellar space can also approach the Solar system from any direction (the stars in the solar 
      neighbourhood are essentially isotropically distributed around the Sun). It may also be argued that comet astrometry can easily be 
      noisy and biased and this may lead to incorrect results (see the different past orbital evolution in Fig.~\ref{inout}). The 
      possibility that bad data may have corrupted the current orbit estimate and produce unreliable formal uncertainties cannot be fully 
      neglected as C/2018~V1 was observed at low solar elongation. However, one has to assume that JPL's SBDB has procedures in place to 
      minimize these issues. In addition to being perhaps a first-time visitor from the Oort Cloud, recently perturbed by a stellar fly-by,
      it can also be argued that C/2018~V1 could have an interstellar origin. 

      De la Fuente Marcos et al. (\citeyear{2018MNRAS.476L...1D}) have shown that the distribution of geocentric equatorial coordinates of 
      the radiants of known hyperbolic minor bodies is not isotropic but probably consistent with the one induced by one or more relatively
      recent stellar passages through the Oort Cloud (see e.g. \citealt{2002A&A...396..283D}). One of such passages was that of the binary 
      star WISE~J072003.20-084651.2, also known as Scholz's star \citep{2015ApJ...800L..17M}. The distribution of radiants of known 
      hyperbolic minor bodies in the sky plotted in figs 3 and 4 of \citet{2018MNRAS.476L...1D} shows that the radiant of C/2018~V1 computed 
      above ---$(13\fh41\pm0\fh02, -48\fdg05\pm0\fdg13)$, if originally unbound--- is well separated from the conspicuous overdensities of 
      radiants present in the figures, but close to the projection of the Galactic disc on the sky, which outlines the Milky Way (the 
      arrival directions of interstellar materials are expected to concentrate towards the Galactic plane, see e.g. 
      \citealt{2004ApJ...600..804M}). These facts suggest that C/2018~V1 could not possibly be a first-time visitor from the Oort Cloud, 
      recently perturbed by a stellar fly-by. The overdensities of radiants identified by \citet{2018MNRAS.476L...1D} could be consistent 
      with the outcome of several stellar passages, but the radiant of C/2018~V1 is very far away from them.  

\begin{figure}
        \centering
         \includegraphics[width=\linewidth]{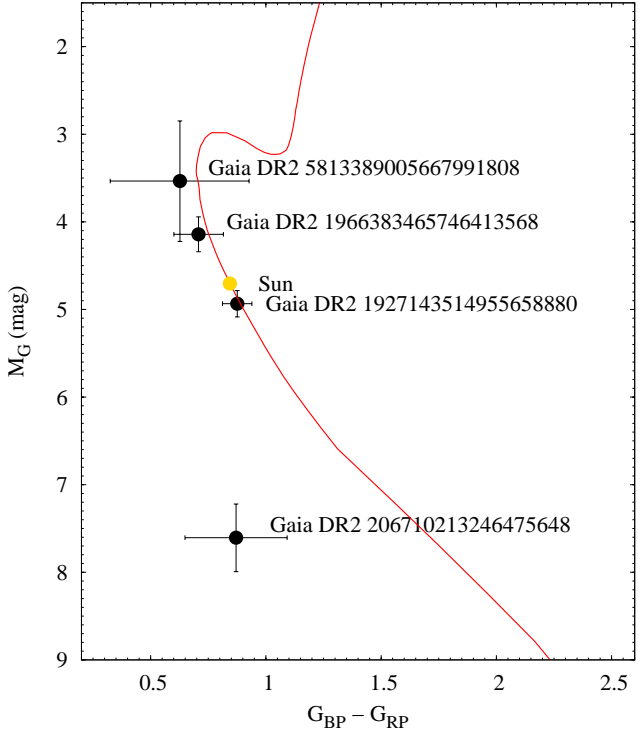}
         \caption{Colour-magnitude diagram for sources in Table~\ref{stars}. One isochrone of age 4.568~Gyr (in red) and solar metallicity 
                  is plotted as a reference, the Sun is plotted as a amber filled circle. See the text for details. 
                 }
         \label{CMD}
      \end{figure}

      The subject of a possible origin of C/2018~V1 within the stars that populate the Galactic neighbourhood of the Sun deserves further 
      consideration. As pointed out in Section~2.2, our sample is made of sources with both line-of-sight extinction and reddening estimates 
      in {\it Gaia} DR2; therefore, we can construct a CMD with the data to check for consistency and Fig.~\ref{CMD} shows the resulting 
      CMD. This CMD has been obtained as the ones in fig.~5 of \citet{2018A&A...616A..10G}, fig.~19 of \citet{2018A&A...616A...8A}, or 
      fig.~3 in \citet{2018MNRAS.481L..64D}. As a reference, one PARSEC v1.2S + COLIBRI S\_35 \citep{2012MNRAS.427..127B,2017ApJ...835...77M,
      2019MNRAS.485.5666P}\footnote{\url{http://stev.oapd.inaf.it/cgi-bin/cmd}} isochrone of age 4.568~Gyr and solar metallicity is also 
      plotted (in red). The assumed value for the age of the Solar system is the one computed by \citet{2010NatGe...3..637B}, 
      4568.2$^{+0.2}_{-0.4}$~Myr. The value of the metallicity of the Sun used to obtain the isochrone is the one calculated by 
      \citet{2017ApJ...839...55V}, {Z}$_{\odot}=0.0196\pm0.0014$. The uncertainties have been estimated using a Monte Carlo sampling 
      approach similar to the one discussed by \citet{2018ApJ...868...25B} and described by \citet{2019A&A...627A.104D}. The positions in 
      the CMD of two of the entries in Table~\ref{stars} ---{\it Gaia} DR2 1927143514955658880 and 1966383465746413568--- appear to be 
      consistent with being robust solar sibling candidates as they are sufficiently close to the theoretical isochrone. The other two 
      kinematic matches ---{\it Gaia} DR2 206710213246475648 and 5813389005667991808--- are unlikely to share age and metallicity with the 
      Sun, and may be the result of trapping in a spiral corotation resonance as described by \citet{2017ApJ...843...48L}, but see below for 
      a more detailed discussion on the reliability of their astrometric solutions.
%
%
%
%

      One may also argue about the quality of the data collected from {\it Gaia} DR2. The astrometric solution of {\it Gaia} DR2 
      206710213246475648 has been computed using 12 visibility periods, the total number of field-of-view (FoV) transits matched to this 
      source was 30, but the astrometric excess noise was 4.110~mas and its significance 6280, which strongly suggests that the source may 
      not be single. In the case of {\it Gaia} DR2 1927143514955658880, the astrometric solution has been computed using 16 visibility 
      periods, with 33 matches and the astrometric excess noise was 0.000~mas. The astrometric solution of {\it Gaia} DR2 
      1966383465746413568 has been computed using 17 visibility periods, the total number of FoV transits matched to this source was 39 and 
      again the astrometric excess noise was 0.000~mas. In the case of {\it Gaia} DR2 5813389005667991808, the astrometric solution has been 
      computed using 16 visibility periods, with 79 matches, however the astrometric excess noise was 1.132~mas and its significance 825, 
      which suggests that the source might not be single. This analysis is consistent with the CMD in Fig.~\ref{CMD}, sources {\it Gaia} DR2 
      1927143514955658880 and 1966383465746413568 appear to be single and astrometrically well-behaved; in addition, they match the 
      isochrone of age 4.568~Gyr quite well. In sharp contrast, sources {\it Gaia} DR2 206710213246475648 and 5813389005667991808 may not be 
      single and their positions in the CMD of Fig.~\ref{CMD} just confirm that their astrometry may not be reliable. If we apply the 
      criteria discussed by \citet{2018A&A...616A...2L} to identify sources with poor astrometric solutions, {\it Gaia} DR2 
      206710213246475648 and 5813389005667991808 emerge as dubious, but {\it Gaia} DR2 1927143514955658880 and 1966383465746413568 are 
      astrometrically well-behaved sources. We consider {\it Gaia} DR2 1927143514955658880 and 1966383465746413568 as {\it bona fide} solar 
      sibling candidates that deserve further study.

      Our analysis shows that both {\it Gaia} DR2 1927143514955658880 and 1966383465746413568 are robust kinematic analogues of C/2018~V1, 
      but this does not necessarily mean that any of them could be the source of the comet studied here. Therefore, we can only state that 
      the predicted past kinematic properties of C/2018~V1 may be consistent with those of some known stars located relatively close 
      ($\sim$300~pc) to the Sun. Arguing for an actual origin implies that numerical simulations including the source star and the 
      interstellar comet place both objects in close proximity at some time in the past. Such calculations have been performed using the 
      approach described in \citet{2018RNAAS...2b..30D} and no positive results in the form of sufficiently close encounters within the last 
      200~Myr have been found. In any case, this section is merely an outline of how to proceed when testing the what-if scenario that 
      motivates the analysis. 

      In addition to passing minor bodies like `Oumuamua, C/1997~P2 (Spacewatch), C/2008~J4 (McNaught), C/2012~S1 (ISON) or 
      C/2018~V1, the Solar system may host a number of captured extrasolar minor bodies \citep{2019ApJ...872L..10S} as it might arguably be 
      the case of comets C/1996~B2 (Hyakutake) \citep{1996Sci...272.1310M,1996Natur.383..606B,1996Natur.383..418I} and 96P/Machholz~1 
      \citep{2007ApJ...664L.119L,2008AJ....136.2204S} or Jupiter's retrograde co-orbital asteroid (514107) 2015~BZ$_{509}$ 
      \citep{2018MNRAS.477L.117N}, although 96P/Machholz~1 might eventually return to deep space \citep{2015MNRAS.446.1867D}. Slightly 
      hyperbolic bodies like C/2018~V1 are primary candidates to be captured in heliocentric orbits; C/1996~B2, 96P/Machholz~1 or 514107 may 
      have reached the Solar system at low relative speed ---below the 0.5~km~s$^{-1}$ critical value pointed out by 
      \citet{1982ApJ...255..307V}--- before being captured. This may have been the case of C/2018~V1 as well. The existence of low relative 
      velocity interstellar interlopers also has strong implications on the effectiveness of the planetary seeding mechanism proposed by 
      \citet{2019ApJ...874L..34P} and further developed by \citet{2019MNRAS.487.3324G}; the presence of large numbers of low-relative-speed 
      ejected planetesimals within an already evolved star-forming region may further accelerate the process of planet formation via 
      captures.

      Regarding the observability and even accessibility of interstellar interlopers (see e.g. the discussion in 
      \citealt{2019ApJ...874L..11E}), those moving at low relative velocities with respect to the Sun are ideal targets not only because
      of their extended visibility windows compared to that of 1I/2017~U1 (`Oumuamua) ---that was 80~d--- but also because they are far 
      easier targets for {\it in situ} exploration (see e.g. \citealt{2018AJ....155..217S}) via fast-response missions. {\it Comet 
      Interceptor},\footnote{\url{http://www.cometinterceptor.space/}} a fast-class mission recently selected by the European Space Agency 
      (ESA),\footnote{\url{http://www.esa.int/Our_Activities/Space_Science/ESA_s_new_mission_to_intercept_a_comet}} aims at visiting one 
      interstellar interloper, starting its journey from the Sun-Earth Lagrange point L$_{2}$. A slow interstellar comet is probably the 
      most feasible target in terms of pre-encounter planning and accessibility for this future ESA's mission. However, Table~\ref{elements} 
      and Fig.~\ref{inout} clearly show that it could be difficult to make the right decision if the quality of the orbit determination is 
      not good enough.  

      On the other hand, objects like C/2018~V1 can be natural probes into the resonant conditions that may surround the space just beyond 
      the Oort Cloud and into the population of low-relative-velocity stars located near the Sun. In this regard, high resolution 
      spectroscopy of the kinematic analogues of C/2018~V1 presented in Table~\ref{stars} can help in confirming or rejecting any connection 
      with the Sun and perhaps C/2018~V1.
  
  \section{Conclusions}
      In this paper, we have studied the pre- and post-encounter orbital evolution of C/2018~V1 (Machholz-Fujikawa-Iwamoto), a slightly 
      hyperbolic comet first observed on 2018 November 7. This research has made use of the latest comet data, $N$-body simulations, 
      {\it Gaia} DR2 data, and statistical analyses. Our conclusions can be summarized as follows.
      \begin{enumerate}[(i)]
         \item We show that C/2018~V1 has little to no dynamical correlation with known parabolic or hyperbolic comets when considering its 
               overall orbital orientation in space.
         \item We confirm that, after analyzing an extensive set of $N$-body simulations, C/2018~V1 may have come from the Oort Cloud but
               it will leave the Solar system aiming for interstellar space after its recent perihelion passage, never to return. It is 
               however not possible to discard an extrasolar origin for this object using only the available data. 
         \item If originally unbound, C/2018~V1 may have entered the Solar system nearly 1~Myr ago at very low relative velocity with 
               respect to the Sun. We have carried out a search for nearby stars in {\it Gaia} DR2 that may have kinematics consistent
               with this scenario; two kinematic analogues of C/2018~V1 have been identified ---{\it Gaia} DR2 1927143514955658880 and 
               1966383465746413568--- and they could be solar sibling candidates.
         \item Our analysis shows that comets coming from interstellar space with relatively low velocities with respect to the Sun may not
               be uncommon. 
      \end{enumerate}
      Spectroscopic studies of C/2018~V1 may have been able to confirm if this comet could have an extrasolar origin by finding, or not, a 
      chemical composition consistent with that of well-studied Solar system materials. Comet C/2018~V1 is already a Southern hemisphere 
      object, and it was well positioned for observations from May to July in 2019, but its apparent visual magnitude will go above 25~mag 
      by the end of its 2020 opposition. At 30~mag, the object will become virtually unobservable by the end of 2023.

  \section*{Acknowledgements}
      We thank the referee for her/his constructive reports and helpful suggestions regarding the presentation of this paper and the 
      discussion of our results, S.~J. Aarseth for providing one of the codes used in this research, J. de Leon, J. Licandro and 
      M. Serra-Ricart for discussions on the nature of hyperbolic minor bodies, and  A.~I. G\'omez de Castro for providing access to 
      computing facilities; RdlFM thanks L. Beitia-Antero for extensive discussions on {\it Gaia} DR2 data. This work was partially 
      supported by the Spanish `Ministerio de Econom\'{\i}a y Competitividad' (MINECO) under grants ESP2015-68908-R and ESP2017-87813-R. In 
      preparation of this paper, we made use of the NASA Astrophysics Data System, the ASTRO-PH e-print server, the MPC data server, and the 
      SIMBAD and VizieR data bases operated at CDS, Strasbourg, France. This work has made use of data from the ESA mission {\it Gaia} 
      (\url{https://www.cosmos.esa.int/gaia}), processed by the {\it Gaia} Data Processing and Analysis Consortium (DPAC, 
      \url{https://www.cosmos.esa.int/web/gaia/dpac/consortium}). Funding for the DPAC has been provided by national institutions, in 
      particular the institutions participating in the {\it Gaia} Multilateral Agreement.

  \bsp
  \label{lastpage}
\end{document}